\documentclass[
 reprint,amsmath,amssymb,
 aps,prl
]{revtex4-1}

\usepackage{graphicx}% Include figure files
\usepackage{dcolumn}% Align table columns on decimal point
\usepackage{bm}% bold math
\usepackage[utf8]{inputenc}
\usepackage[T1]{fontenc}
\usepackage{textcomp}
\usepackage{gensymb}
\usepackage{lipsum}
\usepackage{color}
\usepackage{xr}

\begin{document}

\title{Attraction-enhanced emergence of friction in colloidal matter}
 
\author{ Berend van der Meer$^{1,2}$, Taiki Yanagishima$^{2,3}$, Roel P. A. Dullens$^{1,2}$}
 
\affiliation{$^1$Institute for Molecules and Materials, Radboud University, Heyendaalseweg 135, 6525 AJ, Nijmegen, The Netherlands}

\affiliation{$^2$Department of Chemistry, Physical and Theoretical Chemistry Laboratory, University of Oxford, South Parks Road, Oxford OX1 3QZ, United Kingdom}

\affiliation{$^3$Department of Physics, Graduate School of Science, Kyoto University, Kitashirakawa Oiwake-cho, Sakyo-ku, Kyoto, 606-8502, Japan}

\date{\today}

\begin{abstract}
How frictional effects emerge at the microscopic level in particulate materials remains a challenging question, particularly in systems subject to thermal fluctuations due to the transient nature of interparticle contacts. Here, we directly relate particle-level frictional arrest to local coordination in an attractive colloidal model system. We reveal that the orientational dynamics of particles slows down exponentially with increasing coordination number due to the emergence of frictional interactions, the strength of which can be tuned simply by varying the attraction strength. Using a simple computer simulation model, we uncover how the interparticle interactions govern the formation of frictional contacts between particles. Our results establish quantitative relations between friction, coordination and interparticle interactions. This is a key step towards using interparticle friction to tune the mechanical properties of particulate materials.
\end{abstract}

\maketitle
Over the last decade, a consensus has emerged that interparticle friction plays a crucial role in setting the mechanical and flow properties of concentrated suspensions of particles~\cite{fernandez2013microscopic,seto2013discontinuous,comtet2017pairwise,van2009jamming,mari2019force,singh2020shear,wyart2014discontinuous,guy2018constraint,guy2015towards,ikeda2012unified,mari2015discontinuous,royer2016rheological,james2018interparticle,lin2015hydrodynamic,hsiao2017rheological,hsu2018roughness,hsu2021exploring,muller2023toughening,schroyen2019stress,park2019contact,singh2020shear,pradeep2021jamming,ilhan2022roughness}. For granular systems, the importance of frictional contacts has been well established (see e.g.~\cite{wyart2014discontinuous,singh2020shear,guy2015towards,guy2018constraint}). For instance, computer simulations have demonstrated that the contact number, defined as the average number of frictional contacts per particle, plays a pivotal role in governing the behavior of these non-Brownian suspensions~\cite{van2009jamming,mari2019force,singh2020shear}. Specifically, in the limit of large sliding and rolling friction, the suspension’s viscosity diverges when the contact number exceeds 12/5, leading to jamming at packing fractions as low as $\phi_J \approx 0.36$~\cite{singh2020shear}. Also, the fraction of frictional constraints has been considered~\cite{wyart2014discontinuous,guy2018constraint}. However, a key challenge lies in understanding how particle-level frictional arrest is related to local structure, given that both the fraction of frictional constraints and the contact number are system-averaged quantities.

This challenge is particularly pronounced in the case of colloidal — i.e. Brownian — materials, where the emergence of friction at the particle level becomes even more elusive due to interparticle contacts being subject to thermal fluctuations. This renders such interactions transient in nature thereby adding an extra layer of complexity. Consequently, interparticle friction remains a poorly understood control parameter for engineering colloidal materials~\cite{hu2020particle}, despite its potential to tune material properties~\cite{ikeda2012unified,mari2015discontinuous,guy2015towards,royer2016rheological,james2018interparticle,lin2015hydrodynamic,hsiao2017rheological,hsu2018roughness,hsu2021exploring,schroyen2019stress,park2019contact,pradeep2021jamming,ilhan2022roughness,muller2023toughening}. To date, colloidal experiments have explored the role of interparticle friction by examining the effect of changes in particle surface properties on the \emph{macroscopic} response of these systems~\cite{hsiao2017rheological,hsu2018roughness,schroyen2019stress,hsu2021exploring,muller2023toughening}. Consequently, it remains an open question as to how particle-level frictional effects arise in Brownian systems due to the interplay between thermal fluctuations and the microscopic interactions between particles.

In this work, we establish a quantitative relation between the local environment of a particle, as quantified by its coordination number, and particle-level frictional arrest in an attractive colloidal model system. In particular, we reveal that rolling constraints, due to intermittent frictional contacts between pairs of neighbouring particles, give rise to an exponential slowdown of the orientational dynamics of a particle with increasing coordination number. Importantly, our results show that the amount of interparticle friction between pairs of bonded particles can be controllably tuned by varying the strength of the attractive interactions. This opens up avenues towards using frictional interactions in engineering the properties of colloidal materials.

\begin{figure*}
 \begin{center}
\includegraphics[width=1\textwidth]{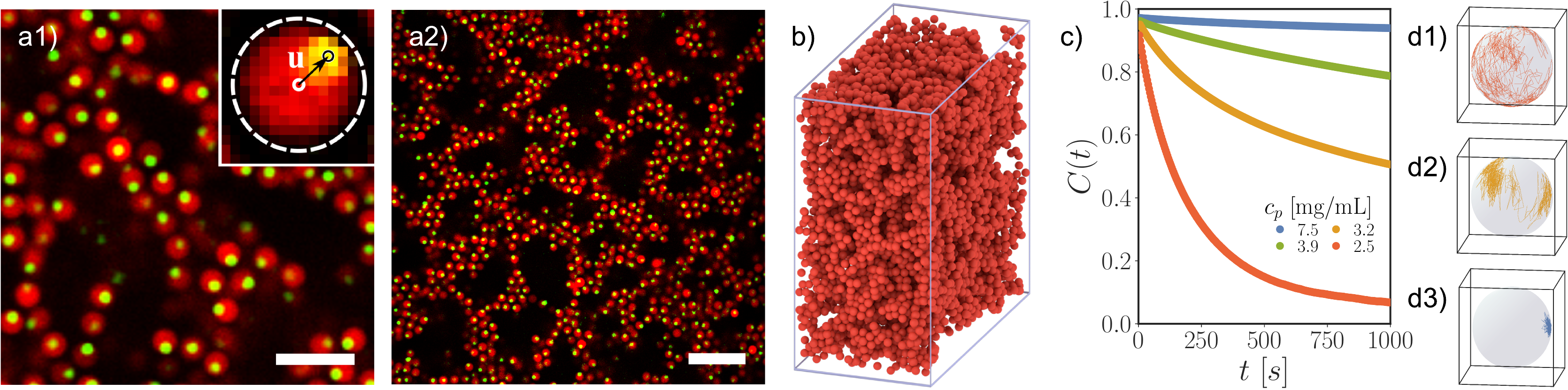} 
\end{center}
 \caption{Orientational dynamics of particles in colloidal gels. (a1, a2) Confocal microscopy images of a colloidal gel of OCULI particles at a polymer concentration $c_p=3.2$ mg/mL. For each particle the centroids of the red "body" and green/yellow "core" are located to determine its 3D orientation vector $\bf u$ (see inset in (a1)). Scale bar: (a1) 10 $\mu$m and (a2) 20$\mu$m. (b) A 3D rendering of all particles within a typical imaging volume of 102 $\times$ 102 $\times$ 50 $\mu$m$^3$. (c) Orientational auto-correlation function $C(t)$  for different polymer concentrations $c_p$. (d1-d3) Typical trajectories of the particle orientation on the unit sphere for polymer concentrations of  2.5, 3.2, and 7.5 mg/mL (top to bottom). (a-d) Colloid volume fraction $\phi\approx 0.3$.}
 \label{fig:fig1}
 \end{figure*}

In our experiments, we use recently developed colloidal particles~\cite{yanagishima2021particle,kamp2021contact}, known as OCULI particles~\cite{yanagishima2021particle}, which have a uniform composition and a non-uniform fluorescence profile [Fig.~\ref{fig:fig1}(a)]. In particular, these OCULI particles feature an off-centre core, labelled with a complementary fluorescent marker, and a non-fluorescent outer layer that enables the simultaneous tracking of both the centroid positions \emph{and the orientations} of all spheres in three dimensions up to particle-particle contact. 
This unique feature allows us to quantitatively study rolling constraints between individual particles due to frictional interactions~\cite{yanagishima2021particle}. Specifically, the particles' orientations rapidly randomise without interparticle contact, whilst the orientational relaxation is significantly slowed down upon making frictional contact. Note that with `frictional contact' we mean any type of interaction that constrains rotational motion due to friction, which can arise from solid-solid friction~\cite{lin2015hydrodynamic,guy2015towards}, but may also be hydrodynamic in nature~\cite{jamali2019alternative,wang2020hydrodynamic}. %Importantly, our method monitors friction via the orientational relaxation, thereby enabling the investigation of rolling friction exclusively \BvdMr{[inaccurate; also torsion]}.}
%Importantly, our method monitors friction through orientational relaxation, which allows us to investigate friction related to rolling and torsion, but not sliding. Additionally, rolling and torsion are effectively superimposed in our measurements.}
Importantly, as friction is monitored through orientational relaxation, this allows for investigation of friction related to rolling and torsion, but not sliding.

% \BvdMr{[I have grouped SM statements; this is OK? A bit late??]} 
We synthesize OCULI particles of diameter $\sigma=3.0 \mu$m following procedures detailed in Ref.~\cite{yanagishima2021particle}. 
%These particles appear smooth under scanning electron microscopy, and hence we infer that the surface roughness must be on the order of tens of nanometres or smaller. 
% CHANGE R1
While these particles appear optically smooth under light microscopy, high-resolution scanning electron microscopy reveals a surface roughness of approximately 50 nm (see SM~\cite{SM}), characterized by a broad distribution of asperity sizes ($\approx 20-100$ nm). 
A short-ranged attractive depletion interaction is introduced between the particles by adding non-adsorbing polystyrene polymers (estimated radius of gyration $R_g\approx 105$nm~\cite{vincent1990calculation}), where the strength of the attractive interactions is varied via the polymer concentration $c_p$~\cite{asakura1954interaction,asakura1958interaction}. 
 These depletion interactions are entropic in nature and arise solely due to excluded-volume effects and  \emph{not} from direct surface-surface interactions, which is why we refer to them as `attractive' interactions rather than `adhesive'. 
 The attractive OCULI particles are suspended in a density- and refractive index-matching organic solvent mixture and imaged using 3D confocal microscopy. More details regarding the colloidal system, attractive interactions, and confocal microscopy experiments are provided in the SM~\cite{SM}. Typical images of the resulting colloidal gel are shown in Fig.~\ref{fig:fig1}(a1, a2) for different length scales, and a 3D reconstruction of the imaged volume is shown in Fig.~\ref{fig:fig1}(b).  These colloidal gels, which are comprised of colloidal particles aggregated into a space-spanning network structure~\cite{dinsmore2002direct,dinsmore2006microscopic,lu2008gelation}, feature rich structural heterogeneity~\cite{van2017linking,whitaker2019colloidal}, making them an ideal system to investigate the relation between local structure and frictional effects at the particle level in colloidal matter.

 \begin{figure*}
\begin{center}
\includegraphics[width=1\textwidth]{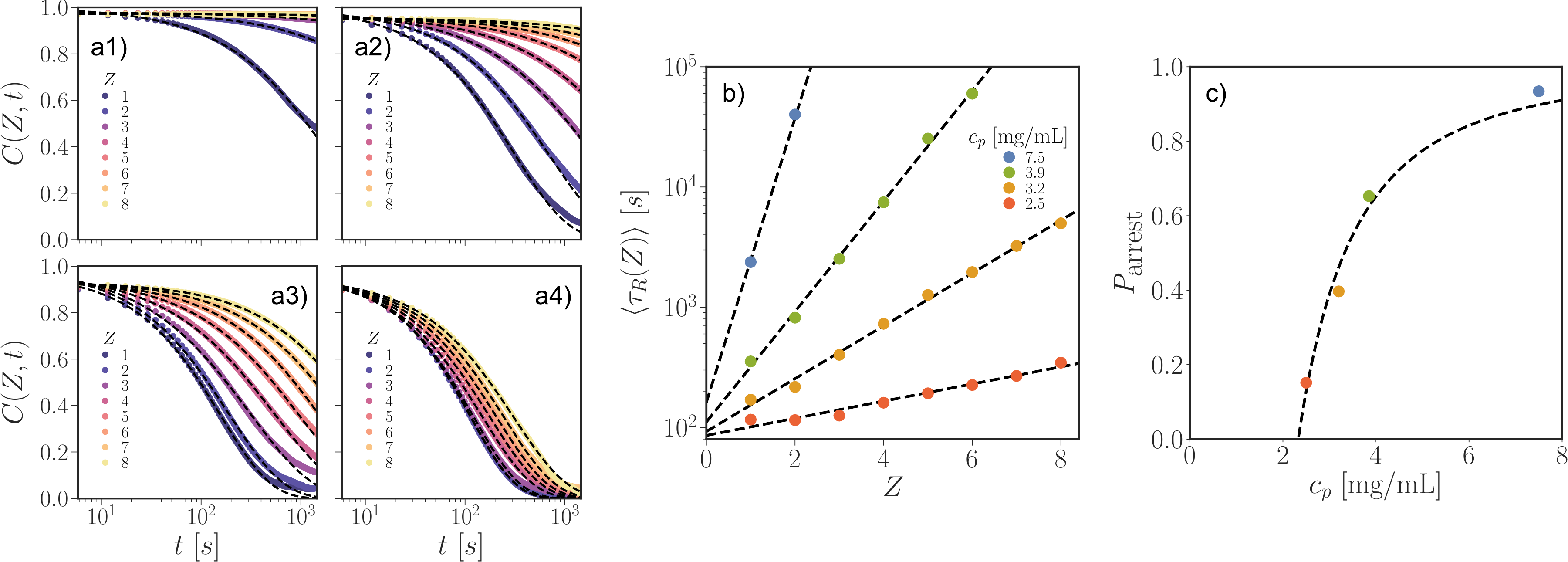} 
\end{center}
\caption{
Coordination-dependent frictional arrest. (a1-a4) The coordination-dependent orientational autocorrelation function $C(Z,t)$ for polymer concentrations $c_p = $ 7.5, 3.9, 3.2, and 2.5 mg/mL, respectively. Dashed lines correspond to stretched-exponential fits. (b) The mean orientational relaxation time $\langle \tau_R(Z) \rangle$ as a function of the coordination number $Z$ for different polymer concentrations $c_p$. Dashed lines are fits to Eq.~\ref{eq:model}. (c) The probability, i.e. fraction of time, that a pair of neighbouring particles is in the `contact' state $P_{\mathrm{arrest}}$ for different polymer concentrations $c_p$. Dashed line is a guide to the eye. (a-c) Colloid volume fraction $\phi\approx 0.3$.}
 \label{fig:fig2}
 \end{figure*}

%\section*{Ensemble-averaged orientational dynamics} 
To investigate the emergence of rolling constraints due to interparticle friction, we monitor  the orientational dynamics of the particles using the orientational auto-correlation function 
\begin{equation}
     C(t)=\langle{\bf u}_i(t) \cdot {\bf u}_i(0) \rangle
\label{eq:autocorr}
\end{equation}
with ${\bf u}_i(t)$ the  unit orientation vector of particle $i$ and $\langle \cdot \cdot \rangle$ denoting an ensemble-average over all particles. Importantly, rolling and torsion are effectively superimposed in these measurements. As shown in Fig.~\ref{fig:fig1}(c), for weak attractive interactions $C(t)$ rapidly decays indicating a fast randomisation of the particle orientation. For stronger attractive interactions, this orientational relaxation is strongly slowed down or even arrested, giving rise to a clear plateau in $C(t)$. This mobile-to-arrested transition of the orientational dynamics is further illustrated by plotting a  typical trajectory of the particle orientation on the unit sphere: at low polymer concentration the particle orientation undergoes constant Brownian rotation [Fig.~\ref{fig:fig1}(d1)], while at higher polymer concentration this motion becomes intermittent [Fig.~\ref{fig:fig1}(d2)] and eventually fully arrested [Fig.~\ref{fig:fig1}(d3)]. 
Our data thus highlight that the emergence of interparticle friction is directly governed by the strength of the attractive interactions. However, the attractive interactions alone cannot explain the observed slowdown in orientational dynamics. This is because the attractive forces act centrally and do not exert tangential constraints, thus lacking the ability to directly hinder rotation. Instead, we attribute the observed arrest to the attractive interactions working in concert with a friction mechanism that naturally occurs in our system (e.g. solid-solid friction~\cite{lin2015hydrodynamic,guy2015towards} and/or hydrodynamic lubrication interactions~\cite{jamali2019alternative,wang2020hydrodynamic}). In other words, the attractive forces promote particles being in close proximity of one another, resulting in a more extensive `sampling' of the tangential frictional interactions between particles, which ultimately leads to the observed `attraction-enhanced' orientational arrest.
Importantly, these rolling constraints do not arise from adhesive surface-surface interactions between the particles. Instead, the origin lies in the enhanced influence of frictional forces caused by the closer proximity of particles due to attraction. We note that previous studies have labelled contacts that restrict rolling in attractive systems as `adhesive contacts'~\cite{richards2020role,richards2021turning,larsen2023rheology}.

%\BvdM{It is important to note that these rolling constraints do not arise directly from an adhesive surface-surface interaction between the particles. The origin of the rolling constraint in our system is not due to adhesion, but rather due to the enhanced influence of frictional forces caused by the closer proximity of particles. Nonetheles, previous studies have labelled contacts that restrict rolling in attractive systems ~\cite{richards2020role,richards2021turning,larsen2023rheology} as `adhesive contacts', }
%\BvdMr{It is important to distinguish this phenomenon from adhesive rolling resistance. While some studies use the term `adhesive contacts' for contacts that restrict rolling in attractive systems ~\cite{richards2020role,richards2021turning,larsen2023rheology}, the origin of the rolling constraint in our system is not due to adhesion, but rather due to the enhanced influence of frictional forces caused by the closer proximity of particles.} %
%\BvdM{[Not sure about i) the term: adhesive rolling resistance --- seems odd ii) of we toch niet beter kunnen zeggen dat attract+fric vaak als  adhesive contact model is gelabelled.. maar dat het niet door een simpele adhesive surface-surface colloidal interaction komt (e.g. tethering of surface-bound DNA/polymers]}

The precise friction mechanism underlying rolling constraints in our system remains elusive as the experimental resolution is insufficient to distinguish between different friction mechanisms. Both solid-solid friction and hydrodynamic lubrication interactions likely contribute, possibly in combination. Crucially, particle surface roughness is an essential factor in inducing rolling constraints, irrespective of the dominant friction mechanism. More specifically, while tangential lubrication interactions, being significantly weaker, can be considered negligible, the normal lubrication forces between surface asperities can fully restrict  tangential motion of the rough particles, effectively giving rise to hydrodynamic frictional contact~\cite{jamali2019alternative,wang2020hydrodynamic}. To investigate the presence of such surface asperities, we conducted high-resolution SEM imaging which revealed a surface roughness of approximately 50 nm, characterized by a broad distribution of asperity sizes ($\approx 20-100$ nm). Although this roughness is modest, it is comparable to that considered theoretically in Refs.~\cite{jamali2019alternative,wang2020hydrodynamic}, and thus supports the idea that hydrodynamic lubrication between asperities may indeed contribute to the frictional arrest observed in our experiments. However, given the close proximity of particles due to depletion forces and the significant surface roughness relative to the depletion interaction range ($\sim200$ nm), both direct solid-solid contact and hydrodynamic friction through asperity interactions are plausible mechanisms. 
%\BvdM{Note that our particles appear smooth under optical microscopy, highlighting that surface roughness is often overlooked as a potential factor in colloidal systems.}
%As colloidal particles (including ours) often appear smooth under optical microscopy, surface roughness is often overlooked as a potential factor in the behaviour of colloidal systems.

%\section*{Coordination-dependent orientational dynamics} 
The ensemble-averaged description provided by $C(t)$ offers no microscopic insight as to how the frictional arrest of a particle arises due to the interaction with neighbouring particles. To establish this quantitative relation between the local environment of a particle and its frictional arrest, we first quantify the local environment of a particle by its time-averaged coordination number $Z_i$ (see SM~\cite{SM}), which is arguably the simplest local structural predictor for orientational arrest. Then, we calculate the orientational autocorrelation function [Eq.~\ref{eq:autocorr}] again, but now for each coordination number separately:
\begin{equation}
    C(Z,t)=\langle{\bf u}_i(t) \cdot {\bf u}_i(0) \rangle_Z
\label{eq:autocorrZ}
\end{equation}
with $Z$ the  coordination number  and $\langle \cdot \cdot \rangle_Z$ denoting an average over all particles with local coordination $Z_i=Z$. We plot $C(Z,t)$ for a range of attraction strengths in Fig.~\ref{fig:fig2}(a1-a4) and observe a clear slowing down of the orientational relaxation with increasing local coordination number, which is especially pronounced at higher attraction strengths [Fig.~\ref{fig:fig2}(a1-a3)]. Next, we define the mean orientational relaxation time $\langle \tau_R(Z) \rangle$, which is a measure for the frictional arrest of the orientational dynamics, as follows~\cite{lindsey1980detailed,alvarez1991relationship}:
\begin{equation}
    \langle \tau_R(Z) \rangle = \int_0^\infty C(Z,t) dt,
\label{eq:mrt}
\end{equation}
where in practice we first fit a stretched exponential to $C(Z,t)$ before doing the integration (see SM~\cite{SM}). In Fig.~\ref{fig:fig2}(b) we plot the mean orientational relaxation time $\langle \tau_R(Z) \rangle$ for different attraction strengths, and $\langle \tau_R(Z) \rangle$ clearly increases exponentially with the local coordination number $Z$. Interestingly, the slope of the exponential slowdown provides a direct measure for how much the orientational dynamics is slowed down per neighbouring particle, which increases strongly with increasing attraction strength.

\begin{figure*}
\begin{center}
\includegraphics[width=0.75\textwidth]{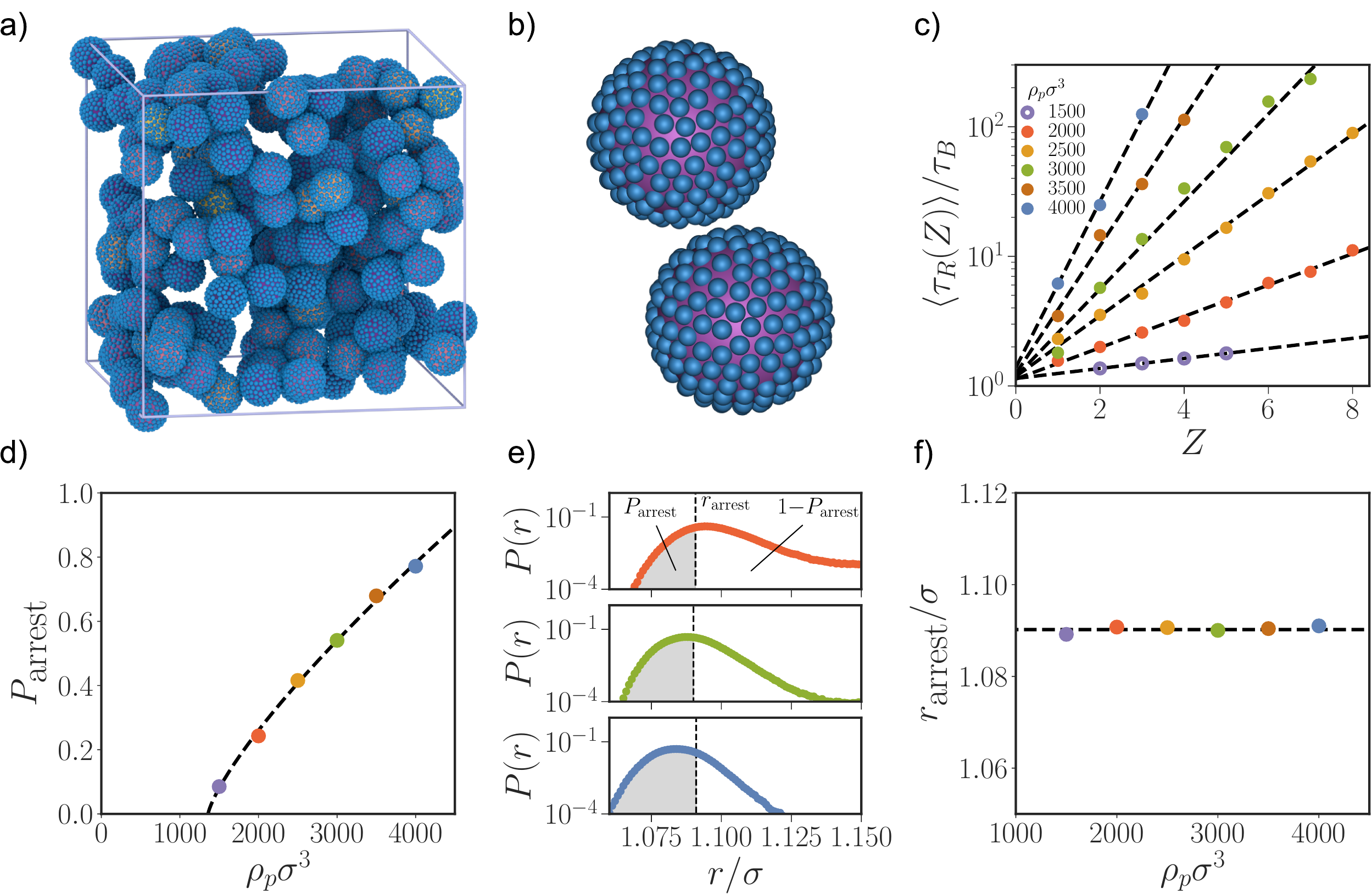}
\end{center}
\caption{Dynamic Monte Carlo simulations of attractive particles with explicit surface roughness. (a) Snapshot of a gel at a dimensionless polymer concentration $\rho_p \sigma^3=3000$. (b) A close-up of the particles showing  clearly the surface topography. (c) The mean orientational relaxation time $\langle \tau_R(Z) \rangle$ as a function of the coordination number $Z$ for different polymer concentrations $\rho_p$. The relaxation times are rendered dimensionless through normalization with the Brownian time $\tau_B=1/2D_r$ with $D_r$ the rotational diffusion constant. Solid and  open markers correspond to gel and fluid phases, respectively. Dashed lines are fits to Eq.~\ref{eq:model}.  (d) The probability of an interparticle bond being frictionally arrested $P_{\mathrm{arrest}}$ as a function of the polymer concentration $\rho_p$. (e) Probability distribution function of bond lengths $P(r)$ for $\rho_p \sigma^3=2000, 3000$ and $4000$ (top to bottom). Here $r_{\mathrm{arrest}}$ (dashed lines) is determined such that $P(r\leq r_{\mathrm{arrest}}) = P_{\mathrm{arrest}}$ (shaded areas), hence separating the `contacting' and `non-contacting' bond-lengths. (f) Extracted $r_{\mathrm{arrest}}$ for different polymer concentrations $\rho_p$, showing no dependence on the attraction strength. (a-f) Colloid number density $\rho\sigma^3=0.38$.}
 \label{fig:fig3}
 \end{figure*}

%\section*{Two-state model for intermittent frictional arrest}
To explain the exponential dependence of this coordination-dependent frictional arrest of the orientational dynamics, we start by considering only two attractive particles. Due to thermal fluctuations the pair of particles is subjected to bond-length fluctuations, which we describe using a two-state picture with fluctuations between `contact' and `no-contact' states. 
In the `contact' state, the particle orientation is arrested due to rolling constraints, resulting in what has been referred to as `adhesive contact' ~\cite{richards2020role,richards2021turning,larsen2023rheology}. 
Conversely, in the `no-contact' state the particle orientation randomises with a relaxation time $\tau_0$.
Note that such a two-state dynamics is reminiscent of the switching between rotationally arrested and diffusive states as observed at high volume fractions \cite{yanagishima2021particle}, where particles are forced together by crowding.
The average orientational relaxation time is then simply proportional to the time spent in the `no-contact' state. This can be expressed as
\begin{equation}
 \langle \tau_R(Z=1) \rangle \approx \frac{\tau_0}{(1-P_{\mathrm{arrest}})},
 \label{eq:modelpair}
 \end{equation}
 where $P_{\mathrm{arrest}}$ is the probability, i.e. fraction of time, that the particle is rotationally arrested due to rolling friction. 
We note that $P_\textrm{arrest}$ is reminiscent of the system-averaged fraction of frictional contacts considered in earlier theoretical work~\cite{wyart2014discontinuous,guy2018constraint}.
 Importantly, the above argument for pairs of particles is easily extended to account for larger coordination numbers by assuming that bond fluctuations between all pairs of neighbouring particles are independent of each other, which leads to the following straightforward extension of Eq.~\ref{eq:modelpair}:
\begin{equation}
    \langle \tau_R(Z) \rangle \approx\frac{\tau_0}{(1-P_\textrm{arrest})^Z}.
    \label{eq:model}
\end{equation}
Here, $(1-P_\textrm{arrest})^Z$ is thus the probability that all bonds of the central particle with its neighbours are in the `no-contact' state. 
Crucially, this predicts the experimentally observed exponential increase of the orientational relaxation time with coordination number (see Fig.~\ref{fig:fig2}(b)), which thus arises due to the fact that a particle is only able to reorient when it is not in frictional contact with any of its neighbours. 
Remarkably, the simple approximation of uncorrelated bond-length fluctuations yields an accurate description of the relaxation behavior at higher coordination numbers.

Within our two-state picture the slope of the exponential slowdown is directly related to  $P_\textrm{arrest}$, the probability of an interparticle bond being in the `contact' state. To quantify this probability we determine $P_{\mathrm{arrest}}$ from our experimental data by fitting the measured relaxation times to Eq.~\ref{eq:model} for each attraction strength (dashed lines in Fig.~\ref{fig:fig2}(b)).  In Fig.~\ref{fig:fig2}(c), we plot $P_{\mathrm{arrest}}$ as a function of the polymer concentration and clearly see that interparticle bonds spend increasingly more time in the `contact' state as the strength of the attractive interactions increases. This implies that the attraction strength directly controls the amount of interparticle friction on the level of pairs of particles.

%\section*{Interparticle interactions and frictional arrest}
To further establish how the interactions govern the formation of interparticle contacts, we employ a simple computer simulation model which also features the competition between Brownian rotation and friction-induced arrest. To this end, we perform dynamic Monte Carlo computer simulations~\cite{sanz2010dynamic,romano2011monte} in which we combine attractive interactions with explicit particle surface roughness (see SM~\cite{SM}), which constrains the particle motion via steric interactions. 
%Change R4
Hydrodynamic interactions are neglected.
The rough particles are modeled as hard, impenetrable particles to which an attractive depletion interaction is added [Fig.~\ref{fig:fig3}(a,b)]. To ensure that sufficiently long time-scales can be explored, the surface roughness is chosen substantially larger than in the experiments.  
%Hence, it is not our goal to directly mimic the experimental situation but much rather to provide a simple model system to test the interplay between attractive interactions and particle-level frictional arrest due to interparticle contacts. 
Hence, it is not our goal to directly mimic the experimental situation, but rather to see whether this simple model framework may be applied to study the interplay between attractive interactions and particle-level frictional arrest.
As shown in Fig.~\ref{fig:fig3}(c), our simple model system captures the behaviour observed in the experiments. Upon plotting $C(Z,t)$ (see SM~\cite{SM}), we observe a clear dependence of the orientational relaxation of particles on their local coordination with a mean relaxation time  $\langle \tau_R(Z) \rangle$ that increases exponentially with particle coordination number $Z$ [Fig.~\ref{fig:fig3}(c)]. Furthermore, we also observe that  $P_{\mathrm{arrest}}$ increases strongly with attraction strength, i.e. interparticle bonding becomes increasingly more frictional [Fig.~\ref{fig:fig3}(d)]. 
%Change R3% 
%The correspondence between the behaviour observed in the simulations and experiments is particularly remarkable considering the large difference in surface roughness of the constituent particles. This underlines the generality of the observed behaviour: the essential physics emerges simply by combining attractive interactions with a constraining `contact' state. 
The striking correspondence between the behaviour observed in the simulations and experiments is remarkable given the significant difference in particle surface roughness and (lack of) hydrodynamics. 
%This indicates that the underlying mechanism driving this behavior is a general phenomenon in Brownian systems, related to the intermittent switching between contact and non-contact states, independent of the specific nature of the frictional forces involved. 
This suggests that the underlying mechanism driving this behavior is a general phenomenon in Brownian systems, arising from the intermittent switching between contact and non-contact states, regardless of the specific nature of the frictional forces involved.
The essential physics emerges simply by combining attractive interactions with a constraining `contact' state, which in a Brownian system allows for transient contact formation.
Our findings emphasize the generality of this behavior and demonstrate the broad applicability of the two-state model in capturing the essential frictional behavior of Brownian systems.

Finally, we reveal that this attraction-enhanced emergence of friction is governed by a length scale associated with the onset of interparticle friction. As it is prohibitively difficult to access such length scale in our experiments due to localization errors and particle polydispersity, we use our simulations to characterize the probability distribution of bond lengths $P(r)$ for a range of attraction strengths by varying the dimensionless polymer concentration $\rho_{p}\sigma^3$ (see SM~\cite{SM}).  Clearly, $P(r)$ gets narrower and shifts to shorter bond lengths with increasing attraction strengths [Fig.~\ref{fig:fig3}(e)]. The length scale associated with frictional arrest, $r_{\mathrm{arrest}}$, now naturally emerges by separating the bond lengths into `contact' and  `no-contact' states via $P(r \leq r_{\mathrm{arrest}})=P_{\mathrm{arrest}}$ (shaded area in Fig.~\ref{fig:fig3}(e)). Interestingly, $r_{\mathrm{arrest}}$ is independent of the attraction strength [Fig.~\ref{fig:fig3}(f)], thus revealing the length scale associated with interparticle contact $r_{\mathrm{arrest}}$ to be a  particle-property, which originates purely from the steric constraints imposed by the surface roughness at bond lengths $r \leq r_{\mathrm{arrest}}$. Our results thus highlight how interparticle frictional interactions can be tuned in colloidal materials: simply by changing the interparticle interactions one can tune the bond-length distribution $P(r)$ to control the fraction of time bonds spend in the `contact' state [shaded area in Fig.~\ref{fig:fig3}(e)].

%\section*{Conclusions}
In summary, we have uncovered how local coordination and interparticle interactions govern particle-level frictional arrest in attractive colloidal matter. Our results provide direct and fundamental insight as to how microscopic frictional interactions can be tuned, which provides a novel avenue towards tailoring macroscopic bulk behaviour in these materials. We note that the existence of frictional interactions in particle gels has remained elusive to date due to lack of a direct experimental measurement. Importantly, almost all theoretical and simulation studies on such systems do not account for interparticle friction. Incorporating frictional constraints to rolling and sliding between particles will likely have a pronounced effect on a variety of phenomena studied in attractive colloidal matter such as gel formation~\cite{jiang2023colloidal}, coarsening and failure~\cite{muller2023toughening}.

%\section{Acknowledgements}

We would like to thank Ellard Hooiveld and Hanne van der Kooij for performing field-emission electron microscopy measurements.
We also thank Joanne Verweij and Jeffrey Urbach for useful discussions. BvdM acknowledges funding from the Netherlands Organisation for Scientific Research (NWO) through a Rubicon grant (Rubicon Grant No. 019.191EN.011)  and Veni grant (Veni Grant No. VI.Veni.212.138). TY acknowledges a Kyoto University Internal Grant for Young Scientists (Start-up). RPAD acknowledges the European Research Council (Consolidator Grant No. 724834-OMCIDC). 
%\section{Author contributions}  
%The project was conceived by all authors. BvdM performed particle synthesis. TY provided general expertise and technical assistance with regards to the synthesis, imaging and tracking of the particles. BvdM performed all experiments, computer simulations and data analysis. All authors discussed and interpreted the results. BvdM and RPAD wrote the manuscript with input from TY. RPAD supervised the project.

%\bibliography{mybib}

\begin{thebibliography}{44}%
\makeatletter
\providecommand \@ifxundefined [1]{%
 \@ifx{#1\undefined}
}%
\providecommand \@ifnum [1]{%
 \ifnum #1\expandafter \@firstoftwo
 \else \expandafter \@secondoftwo
 \fi
}%
\providecommand \@ifx [1]{%
 \ifx #1\expandafter \@firstoftwo
 \else \expandafter \@secondoftwo
 \fi
}%
\providecommand \natexlab [1]{#1}%
\providecommand \enquote  [1]{``#1''}%
\providecommand \bibnamefont  [1]{#1}%
\providecommand \bibfnamefont [1]{#1}%
\providecommand \citenamefont [1]{#1}%
\providecommand \href@noop [0]{\@secondoftwo}%
\providecommand \href [0]{\begingroup \@sanitize@url \@href}%
\providecommand \@href[1]{\@@startlink{#1}\@@href}%
\providecommand \@@href[1]{\endgroup#1\@@endlink}%
\providecommand \@sanitize@url [0]{\catcode `\\12\catcode `\$12\catcode
  `\&12\catcode `\#12\catcode `\^12\catcode `\_12\catcode `\%12\relax}%
\providecommand \@@startlink[1]{}%
\providecommand \@@endlink[0]{}%
\providecommand \url  [0]{\begingroup\@sanitize@url \@url }%
\providecommand \@url [1]{\endgroup\@href {#1}{\urlprefix }}%
\providecommand \urlprefix  [0]{URL }%
\providecommand \Eprint [0]{\href }%
\providecommand \doibase [0]{http://dx.doi.org/}%
\providecommand \selectlanguage [0]{\@gobble}%
\providecommand \bibinfo  [0]{\@secondoftwo}%
\providecommand \bibfield  [0]{\@secondoftwo}%
\providecommand \translation [1]{[#1]}%
\providecommand \BibitemOpen [0]{}%
\providecommand \bibitemStop [0]{}%
\providecommand \bibitemNoStop [0]{.\EOS\space}%
\providecommand \EOS [0]{\spacefactor3000\relax}%
\providecommand \BibitemShut  [1]{\csname bibitem#1\endcsname}%
\let\auto@bib@innerbib\@empty
%</preamble>
\bibitem [{\citenamefont {Fernandez}\ \emph {et~al.}(2013)\citenamefont
  {Fernandez}, \citenamefont {Mani}, \citenamefont {Rinaldi}, \citenamefont
  {Kadau}, \citenamefont {Mosquet}, \citenamefont {Lombois-Burger},
  \citenamefont {Cayer-Barrioz}, \citenamefont {Herrmann}, \citenamefont
  {Spencer},\ and\ \citenamefont {Isa}}]{fernandez2013microscopic}%
  \BibitemOpen
  \bibfield  {author} {\bibinfo {author} {\bibfnamefont {N.}~\bibnamefont
  {Fernandez}}, \bibinfo {author} {\bibfnamefont {R.}~\bibnamefont {Mani}},
  \bibinfo {author} {\bibfnamefont {D.}~\bibnamefont {Rinaldi}}, \bibinfo
  {author} {\bibfnamefont {D.}~\bibnamefont {Kadau}}, \bibinfo {author}
  {\bibfnamefont {M.}~\bibnamefont {Mosquet}}, \bibinfo {author} {\bibfnamefont
  {H.}~\bibnamefont {Lombois-Burger}}, \bibinfo {author} {\bibfnamefont
  {J.}~\bibnamefont {Cayer-Barrioz}}, \bibinfo {author} {\bibfnamefont {H.~J.}\
  \bibnamefont {Herrmann}}, \bibinfo {author} {\bibfnamefont {N.~D.}\
  \bibnamefont {Spencer}}, \ and\ \bibinfo {author} {\bibfnamefont
  {L.}~\bibnamefont {Isa}},\ }\href@noop {} {\bibfield  {journal} {\bibinfo
  {journal} {Physical Review Letters}\ }\textbf {\bibinfo {volume} {111}},\
  \bibinfo {pages} {108301} (\bibinfo {year} {2013})}\BibitemShut {NoStop}%
\bibitem [{\citenamefont {Seto}\ \emph {et~al.}(2013)\citenamefont {Seto},
  \citenamefont {Mari}, \citenamefont {Morris},\ and\ \citenamefont
  {Denn}}]{seto2013discontinuous}%
  \BibitemOpen
  \bibfield  {author} {\bibinfo {author} {\bibfnamefont {R.}~\bibnamefont
  {Seto}}, \bibinfo {author} {\bibfnamefont {R.}~\bibnamefont {Mari}}, \bibinfo
  {author} {\bibfnamefont {J.~F.}\ \bibnamefont {Morris}}, \ and\ \bibinfo
  {author} {\bibfnamefont {M.~M.}\ \bibnamefont {Denn}},\ }\href@noop {}
  {\bibfield  {journal} {\bibinfo  {journal} {Physical Review Letters}\
  }\textbf {\bibinfo {volume} {111}},\ \bibinfo {pages} {218301} (\bibinfo
  {year} {2013})}\BibitemShut {NoStop}%
\bibitem [{\citenamefont {Comtet}\ \emph {et~al.}(2017)\citenamefont {Comtet},
  \citenamefont {Chatt{\'e}}, \citenamefont {Nigu{\`e}s}, \citenamefont
  {Bocquet}, \citenamefont {Siria},\ and\ \citenamefont
  {Colin}}]{comtet2017pairwise}%
  \BibitemOpen
  \bibfield  {author} {\bibinfo {author} {\bibfnamefont {J.}~\bibnamefont
  {Comtet}}, \bibinfo {author} {\bibfnamefont {G.}~\bibnamefont {Chatt{\'e}}},
  \bibinfo {author} {\bibfnamefont {A.}~\bibnamefont {Nigu{\`e}s}}, \bibinfo
  {author} {\bibfnamefont {L.}~\bibnamefont {Bocquet}}, \bibinfo {author}
  {\bibfnamefont {A.}~\bibnamefont {Siria}}, \ and\ \bibinfo {author}
  {\bibfnamefont {A.}~\bibnamefont {Colin}},\ }\href@noop {} {\bibfield
  {journal} {\bibinfo  {journal} {Nature Communications}\ }\textbf {\bibinfo
  {volume} {8}},\ \bibinfo {pages} {1} (\bibinfo {year} {2017})}\BibitemShut
  {NoStop}%
\bibitem [{\citenamefont {van Hecke}(2009)}]{van2009jamming}%
  \BibitemOpen
  \bibfield  {author} {\bibinfo {author} {\bibfnamefont {M.}~\bibnamefont {van
  Hecke}},\ }\href@noop {} {\bibfield  {journal} {\bibinfo  {journal} {Journal
  of Physics: Condensed Matter}\ }\textbf {\bibinfo {volume} {22}},\ \bibinfo
  {pages} {033101} (\bibinfo {year} {2009})}\BibitemShut {NoStop}%
\bibitem [{\citenamefont {Mari}\ and\ \citenamefont
  {Seto}(2019)}]{mari2019force}%
  \BibitemOpen
  \bibfield  {author} {\bibinfo {author} {\bibfnamefont {R.}~\bibnamefont
  {Mari}}\ and\ \bibinfo {author} {\bibfnamefont {R.}~\bibnamefont {Seto}},\
  }\href@noop {} {\bibfield  {journal} {\bibinfo  {journal} {Soft Matter}\
  }\textbf {\bibinfo {volume} {15}},\ \bibinfo {pages} {6650} (\bibinfo {year}
  {2019})}\BibitemShut {NoStop}%
\bibitem [{\citenamefont {Singh}\ \emph {et~al.}(2020)\citenamefont {Singh},
  \citenamefont {Ness}, \citenamefont {Seto}, \citenamefont {de~Pablo},\ and\
  \citenamefont {Jaeger}}]{singh2020shear}%
  \BibitemOpen
  \bibfield  {author} {\bibinfo {author} {\bibfnamefont {A.}~\bibnamefont
  {Singh}}, \bibinfo {author} {\bibfnamefont {C.}~\bibnamefont {Ness}},
  \bibinfo {author} {\bibfnamefont {R.}~\bibnamefont {Seto}}, \bibinfo {author}
  {\bibfnamefont {J.~J.}\ \bibnamefont {de~Pablo}}, \ and\ \bibinfo {author}
  {\bibfnamefont {H.~M.}\ \bibnamefont {Jaeger}},\ }\href@noop {} {\bibfield
  {journal} {\bibinfo  {journal} {Physical Review Letters}\ }\textbf {\bibinfo
  {volume} {124}},\ \bibinfo {pages} {248005} (\bibinfo {year}
  {2020})}\BibitemShut {NoStop}%
\bibitem [{\citenamefont {Wyart}\ and\ \citenamefont
  {Cates}(2014)}]{wyart2014discontinuous}%
  \BibitemOpen
  \bibfield  {author} {\bibinfo {author} {\bibfnamefont {M.}~\bibnamefont
  {Wyart}}\ and\ \bibinfo {author} {\bibfnamefont {M.~E.}\ \bibnamefont
  {Cates}},\ }\href@noop {} {\bibfield  {journal} {\bibinfo  {journal}
  {Physical Review Letters}\ }\textbf {\bibinfo {volume} {112}},\ \bibinfo
  {pages} {098302} (\bibinfo {year} {2014})}\BibitemShut {NoStop}%
\bibitem [{\citenamefont {Guy}\ \emph {et~al.}(2018)\citenamefont {Guy},
  \citenamefont {Richards}, \citenamefont {Hodgson}, \citenamefont {Blanco},\
  and\ \citenamefont {Poon}}]{guy2018constraint}%
  \BibitemOpen
  \bibfield  {author} {\bibinfo {author} {\bibfnamefont {B.~M.}\ \bibnamefont
  {Guy}}, \bibinfo {author} {\bibfnamefont {J.~A.}\ \bibnamefont {Richards}},
  \bibinfo {author} {\bibfnamefont {D.~J.~M.}\ \bibnamefont {Hodgson}},
  \bibinfo {author} {\bibfnamefont {E.}~\bibnamefont {Blanco}}, \ and\ \bibinfo
  {author} {\bibfnamefont {W.~C.~K.}\ \bibnamefont {Poon}},\ }\href@noop {}
  {\bibfield  {journal} {\bibinfo  {journal} {Physical Review Letters}\
  }\textbf {\bibinfo {volume} {121}},\ \bibinfo {pages} {128001} (\bibinfo
  {year} {2018})}\BibitemShut {NoStop}%
\bibitem [{\citenamefont {Guy}\ \emph {et~al.}(2015)\citenamefont {Guy},
  \citenamefont {Hermes},\ and\ \citenamefont {Poon}}]{guy2015towards}%
  \BibitemOpen
  \bibfield  {author} {\bibinfo {author} {\bibfnamefont {B.}~\bibnamefont
  {Guy}}, \bibinfo {author} {\bibfnamefont {M.}~\bibnamefont {Hermes}}, \ and\
  \bibinfo {author} {\bibfnamefont {W.~C.~K.}\ \bibnamefont {Poon}},\
  }\href@noop {} {\bibfield  {journal} {\bibinfo  {journal} {Physical Review
  Letters}\ }\textbf {\bibinfo {volume} {115}},\ \bibinfo {pages} {088304}
  (\bibinfo {year} {2015})}\BibitemShut {NoStop}%
\bibitem [{\citenamefont {Ikeda}\ \emph {et~al.}(2012)\citenamefont {Ikeda},
  \citenamefont {Berthier},\ and\ \citenamefont {Sollich}}]{ikeda2012unified}%
  \BibitemOpen
  \bibfield  {author} {\bibinfo {author} {\bibfnamefont {A.}~\bibnamefont
  {Ikeda}}, \bibinfo {author} {\bibfnamefont {L.}~\bibnamefont {Berthier}}, \
  and\ \bibinfo {author} {\bibfnamefont {P.}~\bibnamefont {Sollich}},\
  }\href@noop {} {\bibfield  {journal} {\bibinfo  {journal} {Physical Review
  Letters}\ }\textbf {\bibinfo {volume} {109}},\ \bibinfo {pages} {018301}
  (\bibinfo {year} {2012})}\BibitemShut {NoStop}%
\bibitem [{\citenamefont {Mari}\ \emph {et~al.}(2015)\citenamefont {Mari},
  \citenamefont {Seto}, \citenamefont {Morris},\ and\ \citenamefont
  {Denn}}]{mari2015discontinuous}%
  \BibitemOpen
  \bibfield  {author} {\bibinfo {author} {\bibfnamefont {R.}~\bibnamefont
  {Mari}}, \bibinfo {author} {\bibfnamefont {R.}~\bibnamefont {Seto}}, \bibinfo
  {author} {\bibfnamefont {J.~F.}\ \bibnamefont {Morris}}, \ and\ \bibinfo
  {author} {\bibfnamefont {M.~M.}\ \bibnamefont {Denn}},\ }\href@noop {}
  {\bibfield  {journal} {\bibinfo  {journal} {Proceedings of the National
  Academy of Sciences}\ }\textbf {\bibinfo {volume} {112}},\ \bibinfo {pages}
  {15326} (\bibinfo {year} {2015})}\BibitemShut {NoStop}%
\bibitem [{\citenamefont {Royer}\ \emph {et~al.}(2016)\citenamefont {Royer},
  \citenamefont {Blair},\ and\ \citenamefont {Hudson}}]{royer2016rheological}%
  \BibitemOpen
  \bibfield  {author} {\bibinfo {author} {\bibfnamefont {J.~R.}\ \bibnamefont
  {Royer}}, \bibinfo {author} {\bibfnamefont {D.~L.}\ \bibnamefont {Blair}}, \
  and\ \bibinfo {author} {\bibfnamefont {S.~D.}\ \bibnamefont {Hudson}},\
  }\href@noop {} {\bibfield  {journal} {\bibinfo  {journal} {Physical Review
  Letters}\ }\textbf {\bibinfo {volume} {116}},\ \bibinfo {pages} {188301}
  (\bibinfo {year} {2016})}\BibitemShut {NoStop}%
\bibitem [{\citenamefont {James}\ \emph {et~al.}(2018)\citenamefont {James},
  \citenamefont {Han}, \citenamefont {de~la Cruz}, \citenamefont {Jureller},\
  and\ \citenamefont {Jaeger}}]{james2018interparticle}%
  \BibitemOpen
  \bibfield  {author} {\bibinfo {author} {\bibfnamefont {N.~M.}\ \bibnamefont
  {James}}, \bibinfo {author} {\bibfnamefont {E.}~\bibnamefont {Han}}, \bibinfo
  {author} {\bibfnamefont {R.~A.~L.}\ \bibnamefont {de~la Cruz}}, \bibinfo
  {author} {\bibfnamefont {J.}~\bibnamefont {Jureller}}, \ and\ \bibinfo
  {author} {\bibfnamefont {H.~M.}\ \bibnamefont {Jaeger}},\ }\href@noop {}
  {\bibfield  {journal} {\bibinfo  {journal} {Nature Materials}\ }\textbf
  {\bibinfo {volume} {17}},\ \bibinfo {pages} {965} (\bibinfo {year}
  {2018})}\BibitemShut {NoStop}%
\bibitem [{\citenamefont {Lin}\ \emph {et~al.}(2015)\citenamefont {Lin},
  \citenamefont {Guy}, \citenamefont {Hermes}, \citenamefont {Ness},
  \citenamefont {Sun}, \citenamefont {Poon},\ and\ \citenamefont
  {Cohen}}]{lin2015hydrodynamic}%
  \BibitemOpen
  \bibfield  {author} {\bibinfo {author} {\bibfnamefont {N.~Y.}\ \bibnamefont
  {Lin}}, \bibinfo {author} {\bibfnamefont {B.~M.}\ \bibnamefont {Guy}},
  \bibinfo {author} {\bibfnamefont {M.}~\bibnamefont {Hermes}}, \bibinfo
  {author} {\bibfnamefont {C.}~\bibnamefont {Ness}}, \bibinfo {author}
  {\bibfnamefont {J.}~\bibnamefont {Sun}}, \bibinfo {author} {\bibfnamefont
  {W.~C.~K.}\ \bibnamefont {Poon}}, \ and\ \bibinfo {author} {\bibfnamefont
  {I.}~\bibnamefont {Cohen}},\ }\href@noop {} {\bibfield  {journal} {\bibinfo
  {journal} {Physical Review Letters}\ }\textbf {\bibinfo {volume} {115}},\
  \bibinfo {pages} {228304} (\bibinfo {year} {2015})}\BibitemShut {NoStop}%
\bibitem [{\citenamefont {Hsiao}\ \emph {et~al.}(2017)\citenamefont {Hsiao},
  \citenamefont {Jamali}, \citenamefont {Glynos}, \citenamefont {Green},
  \citenamefont {Larson},\ and\ \citenamefont
  {Solomon}}]{hsiao2017rheological}%
  \BibitemOpen
  \bibfield  {author} {\bibinfo {author} {\bibfnamefont {L.~C.}\ \bibnamefont
  {Hsiao}}, \bibinfo {author} {\bibfnamefont {S.}~\bibnamefont {Jamali}},
  \bibinfo {author} {\bibfnamefont {E.}~\bibnamefont {Glynos}}, \bibinfo
  {author} {\bibfnamefont {P.~F.}\ \bibnamefont {Green}}, \bibinfo {author}
  {\bibfnamefont {R.~G.}\ \bibnamefont {Larson}}, \ and\ \bibinfo {author}
  {\bibfnamefont {M.~J.}\ \bibnamefont {Solomon}},\ }\href@noop {} {\bibfield
  {journal} {\bibinfo  {journal} {Physical Review Letters}\ }\textbf {\bibinfo
  {volume} {119}},\ \bibinfo {pages} {158001} (\bibinfo {year}
  {2017})}\BibitemShut {NoStop}%
\bibitem [{\citenamefont {Hsu}\ \emph {et~al.}(2018)\citenamefont {Hsu},
  \citenamefont {Ramakrishna}, \citenamefont {Zanini}, \citenamefont
  {Spencer},\ and\ \citenamefont {Isa}}]{hsu2018roughness}%
  \BibitemOpen
  \bibfield  {author} {\bibinfo {author} {\bibfnamefont {C.-P.}\ \bibnamefont
  {Hsu}}, \bibinfo {author} {\bibfnamefont {S.~N.}\ \bibnamefont
  {Ramakrishna}}, \bibinfo {author} {\bibfnamefont {M.}~\bibnamefont {Zanini}},
  \bibinfo {author} {\bibfnamefont {N.~D.}\ \bibnamefont {Spencer}}, \ and\
  \bibinfo {author} {\bibfnamefont {L.}~\bibnamefont {Isa}},\ }\href@noop {}
  {\bibfield  {journal} {\bibinfo  {journal} {Proceedings of the National
  Academy of Sciences}\ }\textbf {\bibinfo {volume} {115}},\ \bibinfo {pages}
  {5117} (\bibinfo {year} {2018})}\BibitemShut {NoStop}%
\bibitem [{\citenamefont {Hsu}\ \emph {et~al.}(2021)\citenamefont {Hsu},
  \citenamefont {Mandal}, \citenamefont {Ramakrishna}, \citenamefont
  {Spencer},\ and\ \citenamefont {Isa}}]{hsu2021exploring}%
  \BibitemOpen
  \bibfield  {author} {\bibinfo {author} {\bibfnamefont {C.-P.}\ \bibnamefont
  {Hsu}}, \bibinfo {author} {\bibfnamefont {J.}~\bibnamefont {Mandal}},
  \bibinfo {author} {\bibfnamefont {S.~N.}\ \bibnamefont {Ramakrishna}},
  \bibinfo {author} {\bibfnamefont {N.~D.}\ \bibnamefont {Spencer}}, \ and\
  \bibinfo {author} {\bibfnamefont {L.}~\bibnamefont {Isa}},\ }\href@noop {}
  {\bibfield  {journal} {\bibinfo  {journal} {Nature Communications}\ }\textbf
  {\bibinfo {volume} {12}},\ \bibinfo {pages} {1} (\bibinfo {year}
  {2021})}\BibitemShut {NoStop}%
\bibitem [{\citenamefont {M{\"u}ller}\ \emph {et~al.}(2023)\citenamefont
  {M{\"u}ller}, \citenamefont {Isa},\ and\ \citenamefont
  {Vermant}}]{muller2023toughening}%
  \BibitemOpen
  \bibfield  {author} {\bibinfo {author} {\bibfnamefont {F.~J.}\ \bibnamefont
  {M{\"u}ller}}, \bibinfo {author} {\bibfnamefont {L.}~\bibnamefont {Isa}}, \
  and\ \bibinfo {author} {\bibfnamefont {J.}~\bibnamefont {Vermant}},\
  }\href@noop {} {\bibfield  {journal} {\bibinfo  {journal} {Nature
  Communications}\ }\textbf {\bibinfo {volume} {14}},\ \bibinfo {pages} {5309}
  (\bibinfo {year} {2023})}\BibitemShut {NoStop}%
\bibitem [{\citenamefont {Schroyen}\ \emph {et~al.}(2019)\citenamefont
  {Schroyen}, \citenamefont {Hsu}, \citenamefont {Isa}, \citenamefont
  {Van~Puyvelde},\ and\ \citenamefont {Vermant}}]{schroyen2019stress}%
  \BibitemOpen
  \bibfield  {author} {\bibinfo {author} {\bibfnamefont {B.}~\bibnamefont
  {Schroyen}}, \bibinfo {author} {\bibfnamefont {C.-P.}\ \bibnamefont {Hsu}},
  \bibinfo {author} {\bibfnamefont {L.}~\bibnamefont {Isa}}, \bibinfo {author}
  {\bibfnamefont {P.}~\bibnamefont {Van~Puyvelde}}, \ and\ \bibinfo {author}
  {\bibfnamefont {J.}~\bibnamefont {Vermant}},\ }\href@noop {} {\bibfield
  {journal} {\bibinfo  {journal} {Physical Review Letters}\ }\textbf {\bibinfo
  {volume} {122}},\ \bibinfo {pages} {218001} (\bibinfo {year}
  {2019})}\BibitemShut {NoStop}%
\bibitem [{\citenamefont {Park}\ \emph {et~al.}(2019)\citenamefont {Park},
  \citenamefont {Rathee}, \citenamefont {Blair},\ and\ \citenamefont
  {Conrad}}]{park2019contact}%
  \BibitemOpen
  \bibfield  {author} {\bibinfo {author} {\bibfnamefont {N.}~\bibnamefont
  {Park}}, \bibinfo {author} {\bibfnamefont {V.}~\bibnamefont {Rathee}},
  \bibinfo {author} {\bibfnamefont {D.~L.}\ \bibnamefont {Blair}}, \ and\
  \bibinfo {author} {\bibfnamefont {J.~C.}\ \bibnamefont {Conrad}},\
  }\href@noop {} {\bibfield  {journal} {\bibinfo  {journal} {Physical Review
  Letters}\ }\textbf {\bibinfo {volume} {122}},\ \bibinfo {pages} {228003}
  (\bibinfo {year} {2019})}\BibitemShut {NoStop}%
\bibitem [{\citenamefont {Pradeep}\ \emph {et~al.}(2021)\citenamefont
  {Pradeep}, \citenamefont {Nabizadeh}, \citenamefont {Jacob}, \citenamefont
  {Jamali},\ and\ \citenamefont {Hsiao}}]{pradeep2021jamming}%
  \BibitemOpen
  \bibfield  {author} {\bibinfo {author} {\bibfnamefont {S.}~\bibnamefont
  {Pradeep}}, \bibinfo {author} {\bibfnamefont {M.}~\bibnamefont {Nabizadeh}},
  \bibinfo {author} {\bibfnamefont {A.~R.}\ \bibnamefont {Jacob}}, \bibinfo
  {author} {\bibfnamefont {S.}~\bibnamefont {Jamali}}, \ and\ \bibinfo {author}
  {\bibfnamefont {L.~C.}\ \bibnamefont {Hsiao}},\ }\href@noop {} {\bibfield
  {journal} {\bibinfo  {journal} {Physical Review Letters}\ }\textbf {\bibinfo
  {volume} {127}},\ \bibinfo {pages} {158002} (\bibinfo {year}
  {2021})}\BibitemShut {NoStop}%
\bibitem [{\citenamefont {Ilhan}\ \emph {et~al.}(2022)\citenamefont {Ilhan},
  \citenamefont {Mugele},\ and\ \citenamefont {Duits}}]{ilhan2022roughness}%
  \BibitemOpen
  \bibfield  {author} {\bibinfo {author} {\bibfnamefont {B.}~\bibnamefont
  {Ilhan}}, \bibinfo {author} {\bibfnamefont {F.}~\bibnamefont {Mugele}}, \
  and\ \bibinfo {author} {\bibfnamefont {M.~H.}\ \bibnamefont {Duits}},\
  }\href@noop {} {\bibfield  {journal} {\bibinfo  {journal} {Journal of Colloid
  and Interface Science}\ }\textbf {\bibinfo {volume} {607}},\ \bibinfo {pages}
  {1709} (\bibinfo {year} {2022})}\BibitemShut {NoStop}%
\bibitem [{\citenamefont {Hu}\ \emph {et~al.}(2020)\citenamefont {Hu},
  \citenamefont {Hsu},\ and\ \citenamefont {Isa}}]{hu2020particle}%
  \BibitemOpen
  \bibfield  {author} {\bibinfo {author} {\bibfnamefont {M.}~\bibnamefont
  {Hu}}, \bibinfo {author} {\bibfnamefont {C.-P.}\ \bibnamefont {Hsu}}, \ and\
  \bibinfo {author} {\bibfnamefont {L.}~\bibnamefont {Isa}},\ }\href@noop {}
  {\bibfield  {journal} {\bibinfo  {journal} {Langmuir}\ }\textbf {\bibinfo
  {volume} {36}},\ \bibinfo {pages} {11171} (\bibinfo {year}
  {2020})}\BibitemShut {NoStop}%
\bibitem [{\citenamefont {Yanagishima}\ \emph {et~al.}(2021)\citenamefont
  {Yanagishima}, \citenamefont {Liu}, \citenamefont {Tanaka},\ and\
  \citenamefont {Dullens}}]{yanagishima2021particle}%
  \BibitemOpen
  \bibfield  {author} {\bibinfo {author} {\bibfnamefont {T.}~\bibnamefont
  {Yanagishima}}, \bibinfo {author} {\bibfnamefont {Y.}~\bibnamefont {Liu}},
  \bibinfo {author} {\bibfnamefont {H.}~\bibnamefont {Tanaka}}, \ and\ \bibinfo
  {author} {\bibfnamefont {R.~P.~A.}\ \bibnamefont {Dullens}},\ }\href@noop {}
  {\bibfield  {journal} {\bibinfo  {journal} {Physical Review X}\ }\textbf
  {\bibinfo {volume} {11}},\ \bibinfo {pages} {021056} (\bibinfo {year}
  {2021})}\BibitemShut {NoStop}%
\bibitem [{\citenamefont {Kamp}\ \emph {et~al.}(2021)\citenamefont {Kamp},
  \citenamefont {de~Nijs}, \citenamefont {Baumberg},\ and\ \citenamefont
  {Scherman}}]{kamp2021contact}%
  \BibitemOpen
  \bibfield  {author} {\bibinfo {author} {\bibfnamefont {M.}~\bibnamefont
  {Kamp}}, \bibinfo {author} {\bibfnamefont {B.}~\bibnamefont {de~Nijs}},
  \bibinfo {author} {\bibfnamefont {J.~J.}\ \bibnamefont {Baumberg}}, \ and\
  \bibinfo {author} {\bibfnamefont {O.~A.}\ \bibnamefont {Scherman}},\
  }\href@noop {} {\bibfield  {journal} {\bibinfo  {journal} {Journal of Colloid
  and Interface Science}\ }\textbf {\bibinfo {volume} {581}},\ \bibinfo {pages}
  {417} (\bibinfo {year} {2021})}\BibitemShut {NoStop}%
\bibitem [{\citenamefont {Jamali}\ and\ \citenamefont
  {Brady}(2019)}]{jamali2019alternative}%
  \BibitemOpen
  \bibfield  {author} {\bibinfo {author} {\bibfnamefont {S.}~\bibnamefont
  {Jamali}}\ and\ \bibinfo {author} {\bibfnamefont {J.~F.}\ \bibnamefont
  {Brady}},\ }\href@noop {} {\bibfield  {journal} {\bibinfo  {journal}
  {Physical Review Letters}\ }\textbf {\bibinfo {volume} {123}},\ \bibinfo
  {pages} {138002} (\bibinfo {year} {2019})}\BibitemShut {NoStop}%
\bibitem [{\citenamefont {Wang}\ \emph {et~al.}(2020)\citenamefont {Wang},
  \citenamefont {Jamali},\ and\ \citenamefont {Brady}}]{wang2020hydrodynamic}%
  \BibitemOpen
  \bibfield  {author} {\bibinfo {author} {\bibfnamefont {M.}~\bibnamefont
  {Wang}}, \bibinfo {author} {\bibfnamefont {S.}~\bibnamefont {Jamali}}, \ and\
  \bibinfo {author} {\bibfnamefont {J.~F.}\ \bibnamefont {Brady}},\ }\href@noop
  {} {\bibfield  {journal} {\bibinfo  {journal} {Journal of Rheology}\ }\textbf
  {\bibinfo {volume} {64}},\ \bibinfo {pages} {379} (\bibinfo {year}
  {2020})}\BibitemShut {NoStop}%
\bibitem [{\citenamefont {Vincent}(1990)}]{vincent1990calculation}%
  \BibitemOpen
  \bibfield  {author} {\bibinfo {author} {\bibfnamefont {B.}~\bibnamefont
  {Vincent}},\ }\href@noop {} {\bibfield  {journal} {\bibinfo  {journal}
  {Colloids and Surfaces}\ }\textbf {\bibinfo {volume} {50}},\ \bibinfo {pages}
  {241} (\bibinfo {year} {1990})}\BibitemShut {NoStop}%
\bibitem [{\citenamefont {Asakura}\ and\ \citenamefont
  {Oosawa}(1954)}]{asakura1954interaction}%
  \BibitemOpen
  \bibfield  {author} {\bibinfo {author} {\bibfnamefont {S.}~\bibnamefont
  {Asakura}}\ and\ \bibinfo {author} {\bibfnamefont {F.}~\bibnamefont
  {Oosawa}},\ }\href@noop {} {\bibfield  {journal} {\bibinfo  {journal} {The
  Journal of Chemical Physics}\ }\textbf {\bibinfo {volume} {22}},\ \bibinfo
  {pages} {1255} (\bibinfo {year} {1954})}\BibitemShut {NoStop}%
\bibitem [{\citenamefont {Asakura}\ and\ \citenamefont
  {Oosawa}(1958)}]{asakura1958interaction}%
  \BibitemOpen
  \bibfield  {author} {\bibinfo {author} {\bibfnamefont {S.}~\bibnamefont
  {Asakura}}\ and\ \bibinfo {author} {\bibfnamefont {F.}~\bibnamefont
  {Oosawa}},\ }\href@noop {} {\bibfield  {journal} {\bibinfo  {journal}
  {Journal of Polymer Science}\ }\textbf {\bibinfo {volume} {33}},\ \bibinfo
  {pages} {183} (\bibinfo {year} {1958})}\BibitemShut {NoStop}%
\bibitem [{SM()}]{SM}%
  \BibitemOpen
  \href@noop {} {\bibinfo  {journal} {See Supplemental Material at [URL will be
  inserted by publisher] for a detailed description of the experiments,
  simulations, and analysis}\ }\BibitemShut {NoStop}%
\bibitem [{\citenamefont {Dinsmore}\ and\ \citenamefont
  {Weitz}(2002)}]{dinsmore2002direct}%
  \BibitemOpen
\bibfield  {journal} {  }\bibfield  {author} {\bibinfo {author} {\bibfnamefont
  {A.~D.}\ \bibnamefont {Dinsmore}}\ and\ \bibinfo {author} {\bibfnamefont
  {D.~A.}\ \bibnamefont {Weitz}},\ }\href@noop {} {\bibfield  {journal}
  {\bibinfo  {journal} {Journal of Physics: Condensed Matter}\ }\textbf
  {\bibinfo {volume} {14}},\ \bibinfo {pages} {7581} (\bibinfo {year}
  {2002})}\BibitemShut {NoStop}%
\bibitem [{\citenamefont {Dinsmore}\ \emph {et~al.}(2006)\citenamefont
  {Dinsmore}, \citenamefont {Prasad}, \citenamefont {Wong},\ and\ \citenamefont
  {Weitz}}]{dinsmore2006microscopic}%
  \BibitemOpen
  \bibfield  {author} {\bibinfo {author} {\bibfnamefont {A.~D.}\ \bibnamefont
  {Dinsmore}}, \bibinfo {author} {\bibfnamefont {V.}~\bibnamefont {Prasad}},
  \bibinfo {author} {\bibfnamefont {I.}~\bibnamefont {Wong}}, \ and\ \bibinfo
  {author} {\bibfnamefont {D.~A.}\ \bibnamefont {Weitz}},\ }\href@noop {}
  {\bibfield  {journal} {\bibinfo  {journal} {Physical Review Letters}\
  }\textbf {\bibinfo {volume} {96}},\ \bibinfo {pages} {185502} (\bibinfo
  {year} {2006})}\BibitemShut {NoStop}%
\bibitem [{\citenamefont {Lu}\ \emph {et~al.}(2008)\citenamefont {Lu},
  \citenamefont {Zaccarelli}, \citenamefont {Ciulla}, \citenamefont
  {Schofield}, \citenamefont {Sciortino},\ and\ \citenamefont
  {Weitz}}]{lu2008gelation}%
  \BibitemOpen
  \bibfield  {author} {\bibinfo {author} {\bibfnamefont {P.~J.}\ \bibnamefont
  {Lu}}, \bibinfo {author} {\bibfnamefont {E.}~\bibnamefont {Zaccarelli}},
  \bibinfo {author} {\bibfnamefont {F.}~\bibnamefont {Ciulla}}, \bibinfo
  {author} {\bibfnamefont {A.~B.}\ \bibnamefont {Schofield}}, \bibinfo {author}
  {\bibfnamefont {F.}~\bibnamefont {Sciortino}}, \ and\ \bibinfo {author}
  {\bibfnamefont {D.~A.}\ \bibnamefont {Weitz}},\ }\href@noop {} {\bibfield
  {journal} {\bibinfo  {journal} {Nature}\ }\textbf {\bibinfo {volume} {453}},\
  \bibinfo {pages} {499} (\bibinfo {year} {2008})}\BibitemShut {NoStop}%
\bibitem [{\citenamefont {van Doorn}\ \emph {et~al.}(2017)\citenamefont {van
  Doorn}, \citenamefont {Bronkhorst}, \citenamefont {Higler}, \citenamefont
  {van~de Laar},\ and\ \citenamefont {Sprakel}}]{van2017linking}%
  \BibitemOpen
  \bibfield  {author} {\bibinfo {author} {\bibfnamefont {J.~M.}\ \bibnamefont
  {van Doorn}}, \bibinfo {author} {\bibfnamefont {J.}~\bibnamefont
  {Bronkhorst}}, \bibinfo {author} {\bibfnamefont {R.}~\bibnamefont {Higler}},
  \bibinfo {author} {\bibfnamefont {T.}~\bibnamefont {van~de Laar}}, \ and\
  \bibinfo {author} {\bibfnamefont {J.}~\bibnamefont {Sprakel}},\ }\href@noop
  {} {\bibfield  {journal} {\bibinfo  {journal} {Physical Review Letters}\
  }\textbf {\bibinfo {volume} {118}},\ \bibinfo {pages} {188001} (\bibinfo
  {year} {2017})}\BibitemShut {NoStop}%
\bibitem [{\citenamefont {Whitaker}\ \emph {et~al.}(2019)\citenamefont
  {Whitaker}, \citenamefont {Varga}, \citenamefont {Hsiao}, \citenamefont
  {Solomon}, \citenamefont {Swan},\ and\ \citenamefont
  {Furst}}]{whitaker2019colloidal}%
  \BibitemOpen
  \bibfield  {author} {\bibinfo {author} {\bibfnamefont {K.~A.}\ \bibnamefont
  {Whitaker}}, \bibinfo {author} {\bibfnamefont {Z.}~\bibnamefont {Varga}},
  \bibinfo {author} {\bibfnamefont {L.~C.}\ \bibnamefont {Hsiao}}, \bibinfo
  {author} {\bibfnamefont {M.~J.}\ \bibnamefont {Solomon}}, \bibinfo {author}
  {\bibfnamefont {J.~W.}\ \bibnamefont {Swan}}, \ and\ \bibinfo {author}
  {\bibfnamefont {E.~M.}\ \bibnamefont {Furst}},\ }\href@noop {} {\bibfield
  {journal} {\bibinfo  {journal} {Nature Communications}\ }\textbf {\bibinfo
  {volume} {10}},\ \bibinfo {pages} {1} (\bibinfo {year} {2019})}\BibitemShut
  {NoStop}%
\bibitem [{\citenamefont {Richards}\ \emph {et~al.}(2020)\citenamefont
  {Richards}, \citenamefont {Guy}, \citenamefont {Blanco}, \citenamefont
  {Hermes}, \citenamefont {Poy},\ and\ \citenamefont
  {Poon}}]{richards2020role}%
  \BibitemOpen
  \bibfield  {author} {\bibinfo {author} {\bibfnamefont {J.~A.}\ \bibnamefont
  {Richards}}, \bibinfo {author} {\bibfnamefont {B.~M.}\ \bibnamefont {Guy}},
  \bibinfo {author} {\bibfnamefont {E.}~\bibnamefont {Blanco}}, \bibinfo
  {author} {\bibfnamefont {M.}~\bibnamefont {Hermes}}, \bibinfo {author}
  {\bibfnamefont {G.}~\bibnamefont {Poy}}, \ and\ \bibinfo {author}
  {\bibfnamefont {W.~C.~K.}\ \bibnamefont {Poon}},\ }\href@noop {} {\bibfield
  {journal} {\bibinfo  {journal} {Journal of Rheology}\ }\textbf {\bibinfo
  {volume} {64}},\ \bibinfo {pages} {405} (\bibinfo {year} {2020})}\BibitemShut
  {NoStop}%
\bibitem [{\citenamefont {Richards}\ \emph {et~al.}(2021)\citenamefont
  {Richards}, \citenamefont {O’Neill},\ and\ \citenamefont
  {Poon}}]{richards2021turning}%
  \BibitemOpen
  \bibfield  {author} {\bibinfo {author} {\bibfnamefont {J.~A.}\ \bibnamefont
  {Richards}}, \bibinfo {author} {\bibfnamefont {R.~E.}\ \bibnamefont
  {O’Neill}}, \ and\ \bibinfo {author} {\bibfnamefont {W.~C.~K.}\
  \bibnamefont {Poon}},\ }\href@noop {} {\bibfield  {journal} {\bibinfo
  {journal} {Rheologica Acta}\ }\textbf {\bibinfo {volume} {60}},\ \bibinfo
  {pages} {97} (\bibinfo {year} {2021})}\BibitemShut {NoStop}%
\bibitem [{\citenamefont {Larsen}\ \emph {et~al.}(2023)\citenamefont {Larsen},
  \citenamefont {S{\o}bye}, \citenamefont {Royer}, \citenamefont {Poon},
  \citenamefont {Larsen}, \citenamefont {Andreasen}, \citenamefont {Drozdov},\
  and\ \citenamefont {Christiansen}}]{larsen2023rheology}%
  \BibitemOpen
  \bibfield  {author} {\bibinfo {author} {\bibfnamefont {T.}~\bibnamefont
  {Larsen}}, \bibinfo {author} {\bibfnamefont {A.~L.}\ \bibnamefont
  {S{\o}bye}}, \bibinfo {author} {\bibfnamefont {J.}~\bibnamefont {Royer}},
  \bibinfo {author} {\bibfnamefont {W.~C.~K.}\ \bibnamefont {Poon}}, \bibinfo
  {author} {\bibfnamefont {T.}~\bibnamefont {Larsen}}, \bibinfo {author}
  {\bibfnamefont {S.~J.}\ \bibnamefont {Andreasen}}, \bibinfo {author}
  {\bibfnamefont {A.~D.}\ \bibnamefont {Drozdov}}, \ and\ \bibinfo {author}
  {\bibfnamefont {J.~D.~C.}\ \bibnamefont {Christiansen}},\ }\href@noop {}
  {\bibfield  {journal} {\bibinfo  {journal} {Journal of Rheology}\ }\textbf
  {\bibinfo {volume} {67}},\ \bibinfo {pages} {81} (\bibinfo {year}
  {2023})}\BibitemShut {NoStop}%
\bibitem [{\citenamefont {Lindsey}\ and\ \citenamefont
  {Patterson}(1980)}]{lindsey1980detailed}%
  \BibitemOpen
  \bibfield  {author} {\bibinfo {author} {\bibfnamefont {C.~P.}\ \bibnamefont
  {Lindsey}}\ and\ \bibinfo {author} {\bibfnamefont {G.~D.}\ \bibnamefont
  {Patterson}},\ }\href@noop {} {\bibfield  {journal} {\bibinfo  {journal} {The
  Journal of Chemical Physics}\ }\textbf {\bibinfo {volume} {73}},\ \bibinfo
  {pages} {3348} (\bibinfo {year} {1980})}\BibitemShut {NoStop}%
\bibitem [{\citenamefont {Alvarez}\ \emph {et~al.}(1991)\citenamefont
  {Alvarez}, \citenamefont {Alegra},\ and\ \citenamefont
  {Colmenero}}]{alvarez1991relationship}%
  \BibitemOpen
  \bibfield  {author} {\bibinfo {author} {\bibfnamefont {F.}~\bibnamefont
  {Alvarez}}, \bibinfo {author} {\bibfnamefont {A.}~\bibnamefont {Alegra}}, \
  and\ \bibinfo {author} {\bibfnamefont {J.}~\bibnamefont {Colmenero}},\
  }\href@noop {} {\bibfield  {journal} {\bibinfo  {journal} {Physical Review
  B}\ }\textbf {\bibinfo {volume} {44}},\ \bibinfo {pages} {7306} (\bibinfo
  {year} {1991})}\BibitemShut {NoStop}%
\bibitem [{\citenamefont {Sanz}\ and\ \citenamefont
  {Marenduzzo}(2010)}]{sanz2010dynamic}%
  \BibitemOpen
  \bibfield  {author} {\bibinfo {author} {\bibfnamefont {E.}~\bibnamefont
  {Sanz}}\ and\ \bibinfo {author} {\bibfnamefont {D.}~\bibnamefont
  {Marenduzzo}},\ }\href@noop {} {\bibfield  {journal} {\bibinfo  {journal}
  {The Journal of Chemical Physics}\ }\textbf {\bibinfo {volume} {132}},\
  \bibinfo {pages} {194102} (\bibinfo {year} {2010})}\BibitemShut {NoStop}%
\bibitem [{\citenamefont {Romano}\ \emph {et~al.}(2011)\citenamefont {Romano},
  \citenamefont {De~Michele}, \citenamefont {Marenduzzo},\ and\ \citenamefont
  {Sanz}}]{romano2011monte}%
  \BibitemOpen
  \bibfield  {author} {\bibinfo {author} {\bibfnamefont {F.}~\bibnamefont
  {Romano}}, \bibinfo {author} {\bibfnamefont {C.}~\bibnamefont {De~Michele}},
  \bibinfo {author} {\bibfnamefont {D.}~\bibnamefont {Marenduzzo}}, \ and\
  \bibinfo {author} {\bibfnamefont {E.}~\bibnamefont {Sanz}},\ }\href@noop {}
  {\bibfield  {journal} {\bibinfo  {journal} {The Journal of Chemical Physics}\
  }\textbf {\bibinfo {volume} {135}},\ \bibinfo {pages} {124106} (\bibinfo
  {year} {2011})}\BibitemShut {NoStop}%
\bibitem [{\citenamefont {Jiang}\ and\ \citenamefont
  {Seto}(2023)}]{jiang2023colloidal}%
  \BibitemOpen
  \bibfield  {author} {\bibinfo {author} {\bibfnamefont {Y.}~\bibnamefont
  {Jiang}}\ and\ \bibinfo {author} {\bibfnamefont {R.}~\bibnamefont {Seto}},\
  }\href@noop {} {\bibfield  {journal} {\bibinfo  {journal} {Nature
  Communications}\ }\textbf {\bibinfo {volume} {14}},\ \bibinfo {pages} {2773}
  (\bibinfo {year} {2023})}\BibitemShut {NoStop}%
\end{thebibliography}

%%%%% BIB

%

\end{document}